\documentclass[twocolumn]{aastex701}

\begin{document}

\title{Massive Perturbers and Transient Pickup Discs in Disc-Crossing Encounters in OJ 287-like Supermassive Black Hole Binaries}

\author[orcid=0000-0002-9597-1015,sname='Garain']{Debojyoti Garain}
\affiliation{Department of Physics and Astronomy, Clemson University, Clemson, SC 29634, USA}
\email[show]{dgarain@clemson.edu}

\author[orcid=0000-0002-1895-6516,sname='Zrake']{Jonathan Zrake}
\affiliation{Department of Physics and Astronomy, Clemson University, Clemson, SC 29634, USA}
\email{jzrake@clemson.edu}

\author[orcid=0000-0003-3633-5403,sname='Haiman']{Zoltan Haiman}
\affiliation{Department of Astronomy, Columbia University, New York, NY 10027, USA}
\affiliation{Department of Physics, Columbia University, New York, NY 10027, USA}
\affiliation{Institute of Science and Technology Austria, AM Campus 1, Klosterneuburg 3400, Austria}
\email{Zoltan.Haiman@ist.ac.at}

\begin{abstract}

We study the hydrodynamical response of a massive black hole accretion disc punctured by a lower mass black hole on an inclined, eccentric orbit, as motivated by the quasi-periodic outbursts seen in the blazar OJ 287. Using three-dimensional smoothed particle hydrodynamics simulations, we explore how the secondary black hole's mass and orbital eccentricity, and the disc thickness and viscosity, affect the time variation of mass accretion onto both black holes. We find that disc-crossing events only lead to significant spikes in the primary black hole's accretion rate when the perturber is quite massive ($q \gtrsim 0.1$) and that such spikes are delayed by roughly a free-fall time ($\sim$ months for OJ 287) after the first disc-crossing. The orbital eccentricity, disc thickness, and viscosity can influence the amplitude and temporal structure of the response; in particular, higher eccentricity causes larger and less delayed delivery of gas to the primary, however the condition $q \gtrsim 0.1$ seems to be robust. We also show that when the secondary is this massive, it generally acquires a ``pickup'' disc which could produce its own luminous signature. For systems like OJ 287, the secondary pickup disc can contain $10^{2-3} M_\odot$, which if accreted over a timescale comparable to the orbital period can power near-Eddington secondary luminosities with thermal emission peaking in the UV/EUV.

\end{abstract}

\keywords{
	\uat{Accretion}{14} ---
	\uat{Active galactic nuclei}{16} ---
	\uat{BL Lacertae objects}{158} ---
	\uat{Galaxy accretion disks}{562} ---
	\uat{Hydrodynamical simulations}{767} ---
	\uat{Supermassive black holes}{1663}
}


\section{Introduction}

Blazars are a class of active galactic nuclei (AGNs) with relativistic jets pointed nearly toward Earth, resulting in strong Doppler boosting and pronounced variability across the electromagnetic spectrum \citep{1995PASP..107..803U}. BL Lacs are highly variable, with changes observed over timescales from decades down to minutes. Such rapid variability implies compact emitting regions in relativistic jets, with the observed timescales shortened by relativistic beaming \citep{1995PASP..107..803U}. OJ 287 is a well-studied BL Lac that is notable for its optical light curve, which has been monitored for over a century \citep{1988ApJ...325..628S, 2013A&A...559A..20H}. Its light curve shows a quasi-periodic, roughly 12-year cycle in brightness with a distinctive double-peaked flare in each cycle. This long-term regularity has been widely interpreted as evidence that OJ 287 hosts a binary supermassive black hole system, in which a secondary massive black hole periodically crosses the primary accretion disc and triggers episodes of enhanced emission \citep{1988ApJ...325..628S}. While the binary hypothesis explains the recurrence pattern of the outbursts, the physical mechanism by which disc-crossing events produce the observed luminosity enhancements remains uncertain.

Several physical mechanisms are proposed by which disc–secondary interactions could produce luminous signatures. One possibility is that shock heating of the primary disc inflates a pressurized bubble, which releases thermal optical radiation when it expands beyond the disc surface and becomes optically thin. This scenario has been developed analytically and explored through hydrodynamical and recent GRMHD simulations \citep{1996ApJ...460..207L,1998ApJ...507..131I,2025ApJ...993L..22R}. Some outbursts seem consistent with such thermal emission, showing low polarization and the absence of strong radio counterparts \citep{2000ApJ...531..744V,2016ApJ...819L..37V}. However, radio observations have found that some flaring episodes show significant polarization and non-thermal spectra, indicating that at least part of the activity is powered by synchrotron emission from the jet \citep{2000ApJ...531..744V,2017Galax...5...83V,2020MNRAS.498L..35K}. Further constraints come from the MOMO monitoring campaign
\citep{2021Univ....7..261K},
which found no evidence for some predicted impact-related thermal flares and showed that low-state optical--UV--X-ray observations indicate a substantially sub-Eddington disc luminosity, suggesting that the primary disc may be too dim or radiatively inefficient to power the pressured-bubble scenario \citep{2023AN....34420126K,2023MNRAS.522L..84K}.

These observations motivate the consideration of other potential energy-release mechanisms, for example, in which disc–secondary interactions lead to modulation of the mass supply to the primary black hole and in turn the jet power. If such jet-dominated flares are linked to binary–disc interaction, then the crossings must do more than generate local shock heating. They must also perturb the disc strongly enough to alter the mass supply reaching the jet-launching region around the primary. This also raises the question of how rapidly the disturbance induced by the crossing propagates inward. If the response is controlled mainly by viscous transport or pressure-wave propagation, the delay could be long and very difficult to predict. Whereas, if the disturbance moves inwards on the free-fall timescale, a delay between the thermal (bubble) and non-thermal (jet) emission components might be reliably derived from the binary orbital parameters. Analytic calculations and idealized simulations suggest that tidal perturbations can indeed redistribute angular momentum driving delayed accretion enhancements \citep{1997ApJ...484..180S}, but a detailed hydrodynamical study of disturbance propagation in discs with parameters appropriate to OJ 287 remains lacking.

A further possibility is that the secondary itself undergoes transient accretion episodes as it captures gas during disc crossings, potentially forming short-lived circum-secondary flows or ``pickup'' disc. Related hydrodynamical studies of embedded perturbers and accreting black hole binaries have shown that strong interactions can lead to gas capture and short-lived secondary accretion structures \citep{2014ApJ...783..134F,2018MNRAS.476.2249T,2019ApJ...879...76B}. If a significant quantity of gas from the primary disk were to be gravitationally captured in orbit around the secondary, then subsequent accretion could power a delayed and higher-frequency emission component.

In this paper, we use three-dimensional smoothed particle hydrodynamics (SPH) simulations to study mass accretion onto both the primary and the secondary massive black holes, resulting from disc-crossing events in a binary with parameters similar to those commonly adopted for OJ 287. We focus on individual disc-crossing passages rather than cumulative evolution across multiple cycles. This isolates the local dynamical response, while leaving open the possibility that persistent non-axisymmetric disturbances from one passage may affect the timing or strength of subsequent flares. Quantifying this cumulative effect requires dedicated multi-passage calculations, which we leave to future work. We present simulations based on fiducial parameters motivated by the binary model of OJ 287 and then extend the study over a wider parameter space in mass ratio, eccentricity, disc thickness, and viscosity. In particular, we focus on three related aspects of the interaction: the efficiency with which tidal disturbances transport gas inward toward the primary, the characteristic delay between disc impact and the resulting accretion enhancement, and the extent to which gas becomes gravitationally bound to the secondary.

We find that triggering significant inward mass flow requires a massive perturber, with the secondary mass generally needing to exceed roughly $10\%$ of the primary mass. Variations in eccentricity, disc thickness, and viscosity modulate the amplitude and temporal structure of the response, but do not set the threshold for enhanced inward transport. The resulting enhancement is delayed relative to the disc crossing, with measured delays comparable to the free-fall time ($\sim$ months for OJ 287) and much shorter than viscous or pressure-wave propagation timescales. We further show that sufficiently massive secondaries can strip and entrain a fraction of the disc material, producing trailing streams that may form transient circum-secondary or ``pickup'' discs. When scaled to OJ 287-like parameters with an Eddington-accreting primary, these pickup discs can reach $10^2$--$10^3\,\rm{M}_\odot$, opening a possible secondary emission channel associated with disc-crossing encounters.

Our paper is organized as follows. In Section~\ref{sec2}, we outline the physical picture of disc–secondary interactions in eccentric, inclined binary systems and describe the key dynamical quantities relevant to disc-crossing events. In Section~\ref{sec3}, we present the numerical methodology, including the SPH setup, initial conditions, and the adopted viscosity prescription. The main results are given in Section~\ref{sec4}, where we analyze the time-dependent accretion response, quantify the delay between disc impact and enhanced accretion, and characterize the formation and evolution of gas bound to the secondary. Finally, in Section~\ref{sec5}, we summarize our main findings and discuss their implications for the interpretation of flaring activity in the OJ 287 system.

\section{Physical picture} \label{sec2}

When a secondary black hole crosses the primary accretion disc on an eccentric, inclined orbit, a fraction of the orbital energy can be transferred to the gas through several distinct but related channels. The most local response is shock heating at the impact site. If the secondary crosses the disc with relative speed $v_{\rm rel}$ through the disc of local scale height $H$, the interaction occurs over a crossing time
\begin{equation}
	t_{\rm cross} \sim \frac{H}{v_{\rm rel}}.
\end{equation}

During this passage, shocks are driven into the surrounding gas, launching hot ejecta above and below the disc plane. In the classical disc-impact picture, this leads to thermal optical emission which arises from the expansion and cooling of this shock-heated material. If the post-impact gas is optically thick, the observed emission depends on photon diffusion through the expanding bubbles. On the other hand, if it is optically thin, the flare duration should more directly follow the local dynamical crossing time. This local thermal channel has already been explored in analytic and numerical studies \citep{1996ApJ...460..207L, 1998ApJ...507..131I}. Our main interest here is instead in the larger-scale response of the disc.

The disc crossing can also excite tidal disturbances that propagate away from the impact site and redistribute angular momentum over a much broader radial range. If these perturbations drive gas inward, they could enhance the accretion rate onto the primary and thereby modulate the power of the relativistic jet. One crucial issue is the timescale of this response. In a viscous picture, the delay is set by
\begin{equation}
	t_{\nu} \sim \frac{R^{2}}{\nu},
\end{equation}
where $R$ is the characteristic radial distance over which the disturbance propagates and $\nu$ is the effective kinematic viscosity. If the disturbance instead propagates through the disc as a pressure wave, the relevant timescale is
\begin{equation}
	t_{\rm s} \sim \frac{R}{c_{\rm s}},
\end{equation}
with $c_{\rm s}$ the sound speed. A third possibility is that the impact places some gas onto strongly eccentric or plunging trajectories, in which case the delay may approach the free-fall time,
\begin{equation}
	t_{\rm ff} \sim \left( \frac{R^{3}}{G M_{1}} \right)^{1/2},
\end{equation}
where $M_{1}$ is the mass of the primary.
Determining which of these timescales best describes the delayed response is crucial for determining whether disc crossings can drive enhanced accretion on the timescales associated with the observed jet-dominated flares in OJ 287.
The efficiency of this inward transport enhancement is ultimately controlled by the strength and duration of the perturbation, with more massive or faster perturbers injecting greater orbital energy into the disc and thereby enhancing non-axisymmetric motions and gas inflow.

A further outcome is transient accretion onto the secondary itself. During a sufficiently strong encounter, some of the disc gas may lose enough energy relative to the secondary to become temporarily bound, forming a trailing stream or a short-lived circum-secondary flow. The natural scale for gas capture is set by the secondary's gravitational sphere of influence, which can be estimated locally by
\begin{equation}
	R_{\rm cap} \cong {\rm{min}}\left(\frac{G M_{2}}{v_{\rm rel}^{2}+c_{\rm s}^{2}}, R_{\rm{Hill}}\right), \label{eq:cap}
\end{equation}
where $M_{2}$ is the secondary mass, and $R_{\rm Hill}=R_{12}(q/3)^{1/3}$ is the Hill radius of the secondary, with $R_{12}$ the instantaneous primary--secondary separation and $q=M_{2}/M_{1}$ the binary mass ratio. The first term in Equation \ref{eq:cap} is the gravitational-focusing, or Bondi--Hoyle-like, capture radius, while the Hill radius accounts for tidal truncation by the primary. The minimum of these two scales therefore provides a conservative estimate of the region within which gas can remain associated with the secondary.
If gas is captured within this scale, it may contribute to a transient ``pickup'' disc and provide an additional channel of delayed emission. The observational importance of this process remains uncertain, but dynamically it provides another channel through which disc crossings redistribute mass and angular momentum within the binary system.

In summary, while local shock heating at the puncture site provides a plausible channel for prompt thermal flare production, our interest here is in the larger-scale hydrodynamical response triggered by the disc crossing. Because we do not follow radiative transfer, we do not attempt to model the detailed thermal emission from the shock-heated gas. Instead, we examine two key dynamical outcomes relevant to OJ 287: delayed accretion onto the primary driven by inward transport of tidally perturbed gas and transient gas capture by the secondary leading to pickup-disc formation. The simulations presented below are therefore designed to quantify the efficiency and timescales of these processes under conditions appropriate to OJ 287.

\section{Code details} \label{sec3}

We perform a suite of 3D SPH simulations using an in-house code that has been extensively validated and applied to a range of astrophysical problems, including tidal disruption events \citep{2023MNRAS.522.4332B,2023JCAP...11..062G,2024ApJ...967..167G,2025MNRAS.542..839G}.
In this work, we model accretion discs around a primary black hole, both in isolation and in the presence of a secondary companion on an eccentric, inclined orbit. We systematically vary the binary mass ratio, orbital eccentricity, disc thickness, and the Shakura–Sunyaev viscosity parameter. Corresponding single black hole simulations with identical disc parameters are performed to provide a controlled reference for the accretion response.
We keep the orbital inclination fixed throughout this parameter survey. Inclination introduces an additional geometric degree of freedom by changing the vertical crossing speed, path length through the disc, and impact geometry, and a systematic exploration of this dependence is therefore left to future work.

We adopt dimensionless code units with $G = M_1 = a = 1$, where $G$ is the gravitational constant, $M_1$ is the primary black hole mass and $a$ is the binary semi-major axis.

\subsection{Disc setup}

Initially the gaseous disc extends from an inner radius $R_{\rm in} = 0.05\,a$ to an outer radius $R_{\rm out} = 2.0\,a$. Particles are distributed using a Monte Carlo sampling technique, consistent with an axisymmetric surface density profile
\begin{equation}
	\Sigma(R) = \Sigma_0 R^{-p},
\end{equation}
with $p = 1$. The normalization is chosen such that the total disc mass is $M_{\rm disc} = 0.006\,M_1$. The initial vertical position of each particle is drawn from a Gaussian distribution with standard deviation $H = c_s / \Omega$, where $c_s$ is the local sound speed and $\Omega$ is the Keplerian angular frequency around the primary. This corresponds to vertical hydrostatic equilibrium for a geometrically thin disc in the gravitational potential of the primary \citep{1973A&A....24..337S}.

The disc is resolved using $N_{\rm part} \approx 2 \times 10^{6}$ SPH particles. Gas self-gravity is neglected, allowing us to focus on the response of a low-mass disc to an external perturbation.

\subsection{Equation of state and disc thickness}

We adopt a locally isothermal equation of state, in which the gas pressure is related to the density through
\begin{equation}
	P = c_s^2 \rho ,
\end{equation}
where the sound speed is prescribed as a function of cylindrical radius and held fixed in time. The sound-speed profile is chosen such that the disc has a constant aspect ratio $H/R$ and is given by
\begin{equation}
	c_s(R) = \frac{H}{R} \left( \frac{G M_1}{R} \right)^{1/2}. \label{eq:cs}
\end{equation}
This corresponds to $H=c_s/\Omega_{\rm K}$ for a Keplerian disc around the primary, with $\Omega_{\rm K}=(GM_1/R^3)^{1/2}$.

The locally isothermal prescription suppresses explicit thermal evolution and allows the disc thickness to remain controlled throughout the simulation. This is a major simplification for the local thermal response to disc impact because shock-generated heat is not evolved, our simulations cannot self-consistently follow the formation, expansion, or radiative cooling of overpressured plumes produced at the puncture site. However, this limitation is less restrictive for the purposes of this work, which focuses on the disc's global dynamical response.

\subsection{Initial velocity field}

The initial azimuthal velocities are set to centrifugal balance including the contribution from the radial pressure gradient,
\begin{equation}
	v_i^2 = \frac{G M_1}{R_i} - c_{s,i}^2 \left(p + 2\right),
\end{equation}
where $R_i$ is the cylindrical distance of the $i$-th particle from the primary and $c_{s,i}$ is the sound speed of the $i$-th particle. Radial and vertical velocities are initially set to zero.

\subsection{Viscosity prescription}

In SPH, an artificial viscosity term is required to capture shocks and to prevent particle interpenetration in converging flows. However, when used as the sole source of dissipation, artificial viscosity is known to introduce excessive and resolution-dependent angular momentum transport in shear-dominated flows such as accretion discs. To minimize its impact on the secular evolution of the disc, we use artificial viscosity only at the minimum level required for numerical stability. Specifically, we adopt the time-dependent artificial viscosity prescription of \cite{1997JCoPh.136...41M}, in which the viscosity parameter $\alpha_{\rm AV}$ varies between $\alpha_{\rm AV,min}=0.1$ and $\alpha_{\rm AV,max}=0.5$, while $\beta_{\rm AV}=2.0$ is held fixed. This choice is motivated by previous disc–binary simulations (e.g. \cite{2016MNRAS.460.1243R}), and ensures that artificial viscosity is enhanced primarily in regions of compression while remaining strongly suppressed in smooth shear flows. Test simulations in which artificial viscosity alone was used to model angular momentum transport resulted in artificially increased accretion rates onto the sinks, motivating the adoption of an explicit physical viscosity prescription.

Physical viscosity is implemented following the SPH formulation of the Navier--Stokes equations introduced by \cite{1994ApJ...431..754F} and described in detail by \cite{2010MNRAS.405.1212L}.
We set the bulk viscosity to zero and compute the kinematic shear viscosity using the dimensionless Shakura--Sunyaev viscosity parameter $\alpha$ \citep{1973A&A....24..337S}. In this prescription, the kinematic viscosity is given by
\begin{equation}
	\nu = \alpha c_s \frac{H}{R} R_1,
\end{equation}
where $R_1$ is the cylindrical distance from the primary black hole and $c_s$ is obtained by Equation \ref{eq:cs}.

\subsection{Gravitational potential and sink particles} \label{sec:sinks}

The black holes are represented by sink particles. The primary mass $M_1$ is present from the beginning of the simulation, while the secondary mass $M_2 = q\,M_1$ is introduced at a later stage. Gas particles evolve under the gravitational influence of the sinks, together with pressure and viscous forces.

Each sink acts as a Newtonian point mass and exerts on the $i$-th gas particle an acceleration
\begin{equation}
	\mathbf{f}_{i,\rm pot} =
	\frac{G M_n (\mathbf{r}_n - \mathbf{r}_i)}
	{|\mathbf{r}_n - \mathbf{r}_i|^3},
\end{equation}
where $n = 1,2$ labels the primary and secondary, and $\mathbf{r}_n$ denotes the position of the sink particle of mass $M_n$. No gravitational softening is applied.

Gas particles are removed once they cross a sink radius $r_{\rm sink}$. The primary sink radius is set to $r_{\rm sink} = 0.05\,a$, coinciding with the inner disc edge, while the secondary sink radius is set to $r_{\rm sink} = 10^{-3}\,a$ to suppress accretion onto the secondary and ensure that it acts primarily as a gravitational perturber. Back-reaction of the gas onto the sinks is neglected.

\subsection{Relaxation and binary configuration}

To minimize numerical transients associated with the initial particle distribution, each disc is first evolved in isolation with only the primary black hole present. During this phase, the disc adjusts self-consistently to the imposed equation of state and viscosity prescription.

We perform a set of simulations with different disc thicknesses and viscosity parameters, and evolve each model for three viscous timescales evaluated at the inner edge,
\begin{equation}
	t_{\nu}(R_{\rm in}) = \frac{2}{3} \frac{R_{\rm in}^2}{\nu(R_{\rm in})},
\end{equation}
ensuring that the inner disc reaches a quasi-steady state. In all cases, the mass accretion rate onto the primary relaxes from its initial transient level and converges to a steady value. We denote this baseline accretion rate for the isolated disc as $\dot{M}_0$.

Following this relaxation phase, the secondary black hole is introduced. The subsequent evolution therefore studies the response of a quasi-steady accretion disc to the perturbation induced by the companion.

The secondary is initialized on a Keplerian orbit around the primary with a prescribed eccentricity. The orbital inclination relative to the disc midplane is fixed at $i=40^\circ$, and the orbital orientation is fixed by choosing the argument of periapsis $\omega=90^\circ$ and longitude of the ascending node $\Omega=0^\circ$.
The secondary is initially placed at apocenter below the disc midplane, and its subsequent evolution is determined solely by the gravitational potential of the primary.

\section{Results} \label{sec4}

\subsection{Simulation details}

We perform a suite of SPH simulations to study the dynamical response of an accretion disc to perturbations induced by a secondary black hole on an  inclined eccentric orbit. To provide a controlled baseline, we define a fiducial model motivated by the OJ 287 system.

The fiducial binary has mass ratio $q = 0.0082$, consistent with observational constraint \citep{2018ApJ...866...11D}, and the orbital eccentricity is chosen to be $e = 0.7$, representative of values inferred from timing models. The disc viscosity parameter is set to $\alpha = 0.26$ \citep{2019ApJ...882...88V}, and the aspect ratio is $H/R = 0.05$, corresponding to a geometrically thin accretion flow. We neglect the spin of the primary black hole, as our focus is on the hydrodynamical response of the disc to individual disc-crossing events rather than on long-term relativistic precession of gas orbits.

Starting from this fiducial configuration, we perform a sequence of controlled simulations in which individual parameters are varied while all others are held fixed. This approach is motivated by two considerations. First, although the adopted fiducial parameters are inspired by existing constraints for OJ 287, significant uncertainties remain in both the binary properties and the disc structure, and it is not a priori clear whether the fiducial model alone can reproduce the observed accretion variability. Second, disc–secondary interactions involve multiple coupled physical processes whose relative importance depends sensitively on the system parameters. We therefore explore a broader parameter space both to test the robustness of the fiducial model and to isolate the physical mechanisms governing the interaction. In particular, the binary mass ratio is varied to determine the threshold at which tidal forcing becomes strong enough to significantly enhance inward mass transport. The orbital eccentricity is varied to examine how the impulsiveness of the disc crossing affects the timing and magnitude of the accretion response. The disc aspect ratio is varied to probe the role of pressure support and vertical structure in setting the propagation speed and damping of large-scale disturbances. Finally, the viscosity parameter is varied to assess how angular momentum transport regulates the conversion of perturbations into accretion onto the primary.

\begin{table}[htbp]
	\centering
	\caption{Simulation parameters}
	\label{tab:sim_params}
	\begin{tabular*}{\columnwidth}{@{\extracolsep{\fill}}lcccc}
		\hline
		Name  & $H/R$ & $\alpha$ & $e$ & $q$ \\
		\hline
		\texttt{sim1} (fiducial OJ 287-like) & 0.05  & 0.26 & 0.7 & 0.0082 \\
		\texttt{sim2}  & 0.05  & 0.26 & 0.7 & 0.025  \\
		\texttt{sim3}  & 0.05  & 0.26 & 0.7 & 0.10   \\
		\texttt{sim4}  & 0.05  & 0.26 & 0.8 & 0.10   \\
		\texttt{sim5}  & 0.05  & 0.26 & 0.9 & 0.10   \\
		\texttt{sim6}  & 0.10  & 0.26 & 0.7 & 0.10   \\
		\texttt{sim7} & 0.05  & 0.10 & 0.7 & 0.10   \\
		\texttt{sim8} & 0.05  & 0.15 & 0.7 & 0.10   \\
		\hline
	\end{tabular*}
\end{table}

This parameter study therefore serves a dual purpose: it allows us to evaluate whether OJ 287-like parameters can account for the accretion disk responses that are suggested by observations, and more generally to characterize the hydrodynamical response of accretion discs to disc-crossing events across a physically relevant range of binary and disc conditions. The full set of simulations and their parameters are summarized in Table~\ref{tab:sim_params}. The explored mass ratios span the range $q = 0.0082$--$0.10$, covering both weak and strong perturbative regimes. Orbital eccentricities in the range $e = 0.7$--$0.9$ are considered to assess the role of increasingly impulsive pericentre passages. The disc aspect ratio is varied between $H/R = 0.05$ and $0.10$ to probe the influence of disc thickness on wave propagation and dissipation, while the viscosity parameter is varied between $\alpha = 0.10$ and $0.26$.

\subsection{Morphological evolution during disc crossing}

We begin by describing the hydrodynamical response of the disc to the passage of the secondary using the fiducial model (\texttt{sim1} in Table~\ref{tab:sim_params}). The secondary is initialized at apocentre and approaches the disc from below the midplane. In code units, the binary orbital period is $P_0 = 2\pi$, and the evolution is therefore conveniently expressed in units of $P_0$.
During a single orbit, the inclined orbit intersects the disc midplane twice. In the adopted orientation, the first mid-plane crossing occurs at $t \simeq 0.45\,P_0$, when the secondary moves from below to above the disc, while the second mid-plane crossing occurs at $t \simeq 0.57\,P_0$, when it moves from above to below the disc.

\begin{figure*}[t!]
	\centering
	\includegraphics[width=1.0\textwidth]{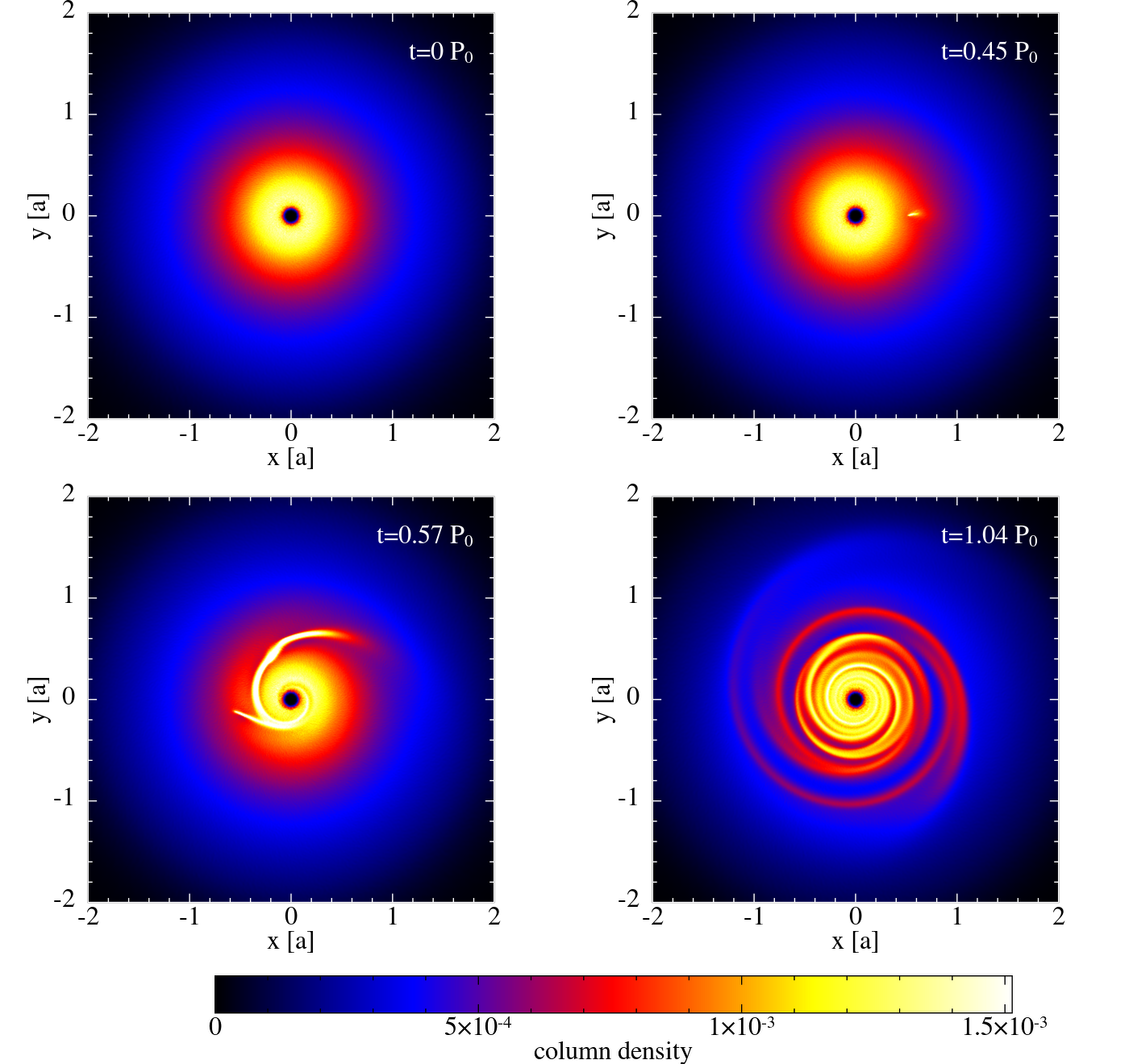}
	\caption{Column density maps showing the morphological evolution of the disc during a single orbital passage of the secondary in the fiducial model. The panels show (top left) the relaxed, axisymmetric disc at $t=0$; (top right) the first mid-plane crossing at $t \simeq 0.45P_0$, when the secondary moves from below to above the disc; (bottom left) the second mid-plane crossing at $t \simeq 0.57P_0$, when the secondary moves from above to below the disc, characterized by prominent spiral density waves; and (bottom right) the subsequent propagation of spiral waves at $t \simeq 1.04\,P_0$. These figures are generated using SPLASH \citep{2007PASA...24..159P}.
		\label{fig:morph}}
\end{figure*}

\begin{figure*}[t!]
	\centering
	\includegraphics[width=1.0\textwidth]{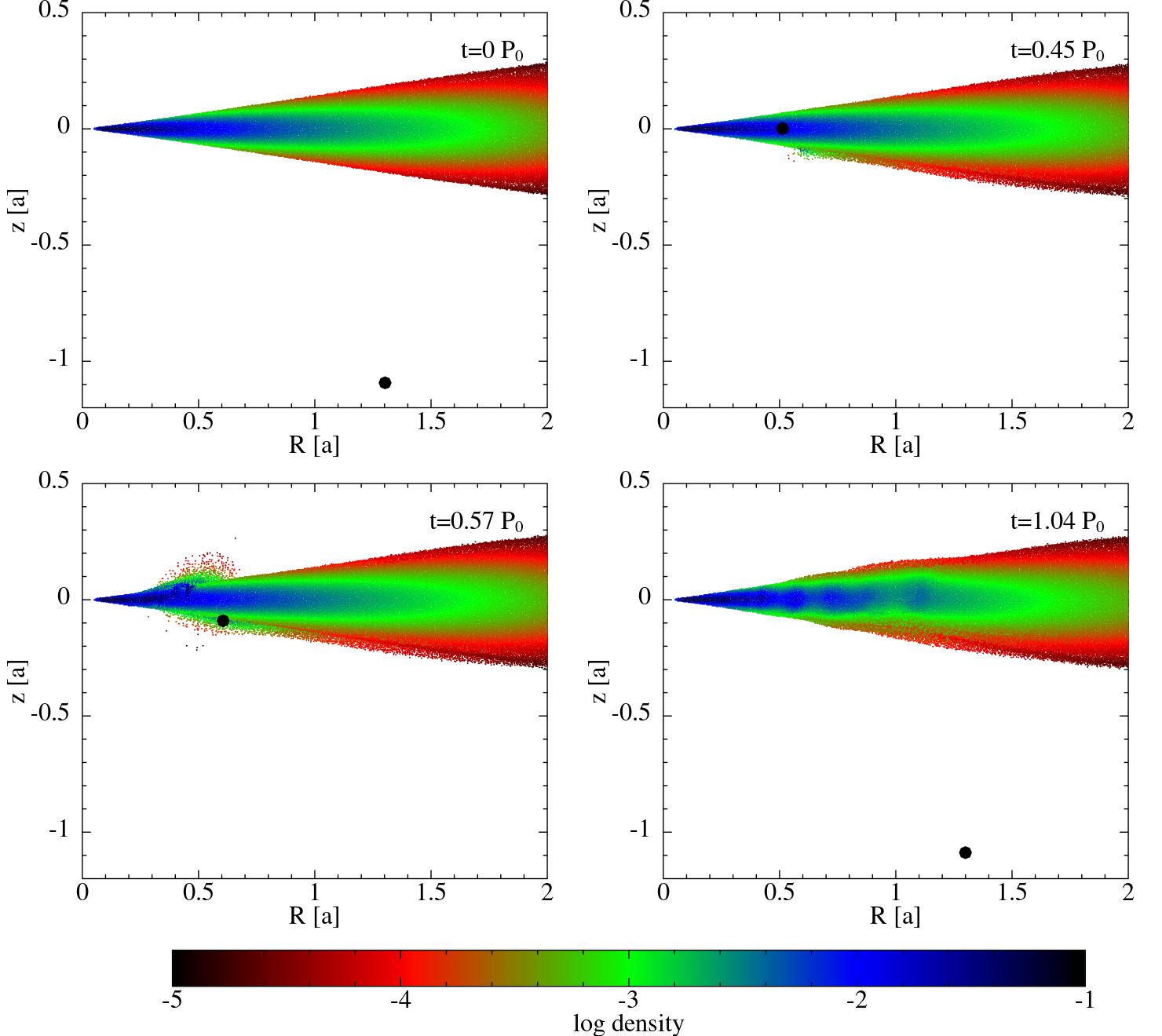}
	\caption{Cylindrical $(R,z)$ particle distributions for the fiducial model, where each point represents an SPH particle colored by the logarithm of its density, and the black dot marks the position of the secondary black hole. The panels correspond to (top left) the relaxed disc at $t=0$; (top right) the first disc mid-plane crossing at $t \simeq 0.45\,P_0$, as the secondary passes from below to above the disc; (bottom left) the second mid-plane crossing at $t \simeq 0.57P_0$, as the secondary passes from above to below the disc, showing strong vertical splashing; and (bottom right) the subsequent evolution at $t \simeq 1.04\,P_0$, where displaced material begins to settle while perturbations propagate radially. These figures are generated using SPLASH.
		\label{fig:RZ}}
\end{figure*}

Figure~\ref{fig:morph} shows face-on column density snapshots at representative times. Prior to the encounter ($t=0$), the disc is axisymmetric and in a quasi-steady state following the relaxation phase.
The first mid-plane crossing produces a localized perturbation at the impact site. By the time of the second mid-plane crossing, the disturbance has strengthened and excites pronounced non-axisymmetric features.
Gas is impulsively displaced above and below the disc surfaces, forming a splash-like structure. This impulsive forcing launches spiral density waves that shear under differential rotation, producing a tightly wound spiral pattern shortly after the second crossing. These spiral features propagate away from the impact region and redistribute mass and angular momentum throughout the disc.
At later times, the initially localized disturbance has spread into a series of concentric, ring-like density enhancements, although residual non-axisymmetric structure persists until the end of the simulation at $t \simeq \,P_0$. Later mid-plane crossings therefore occur in an already perturbed disc, leading to a qualitatively different response compared to the first passage.
The three-dimensional nature of the interaction is illustrated in Figure~\ref{fig:RZ}, which shows $(R,z)$ density slices at the same epochs. These show significant vertical displacement of the gas during disc crossings, including the formation of plumes extending above and below the disc surfaces, followed by fallback. Although the interaction is inherently three-dimensional, the subsequent evolution is dominated by in-plane spiral density waves.

\begin{figure*}[t!]
	\centering
	\includegraphics[width=0.99\textwidth]{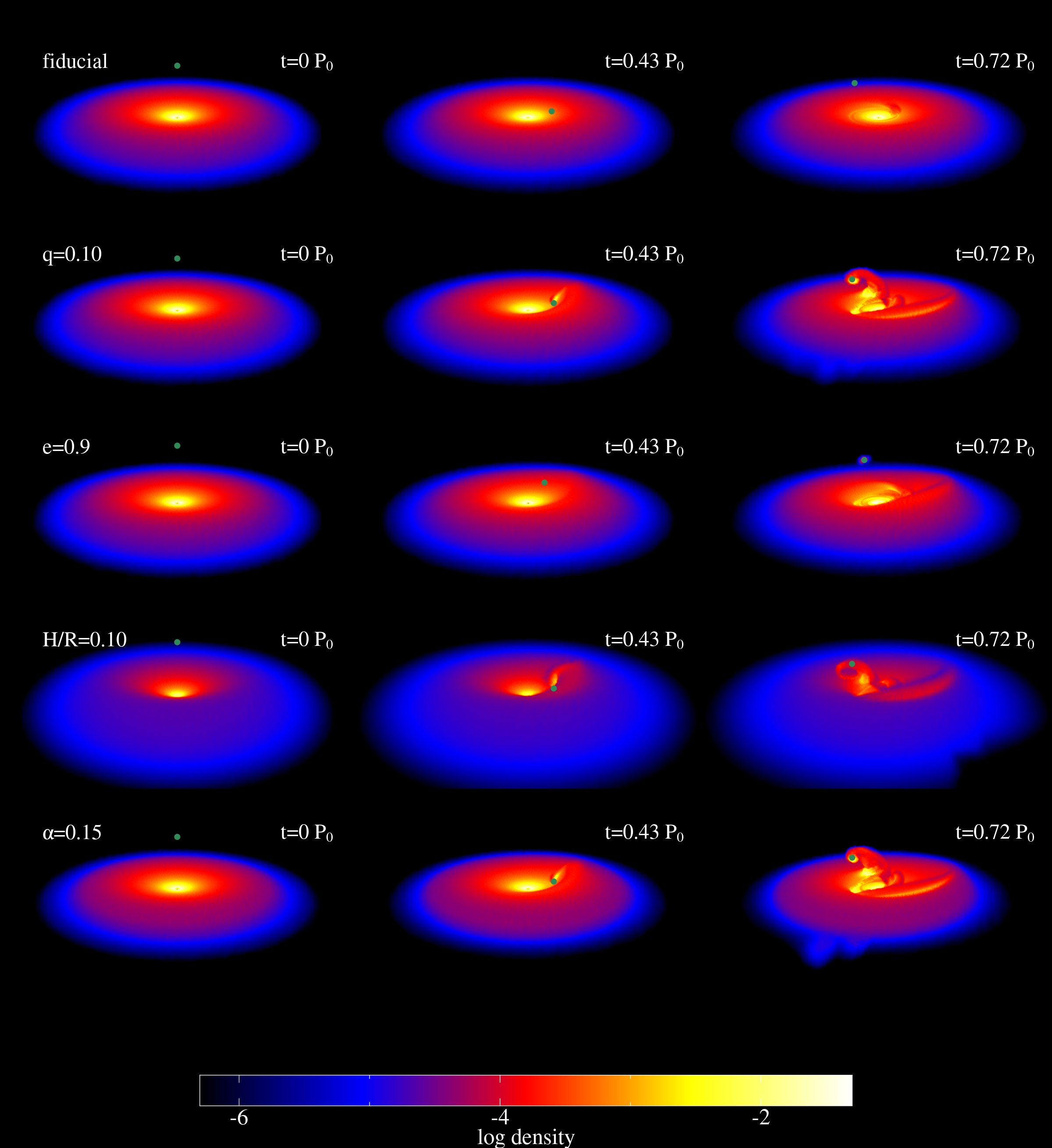}
	\caption{Face-on column density maps showing the dependence of the disc morphology on binary and disc parameters. The first row shows the fiducial model (\texttt{sim1}; $q=0.0082$, $e=0.7$, $H/R=0.05$, $\alpha=0.26$), while each subsequent row changes one parameter relative to the fiducial case: high-mass-ratio model (\texttt{sim3}; $q=0.1$), high-eccentricity model (\texttt{sim5}; $e=0.9$), thicker-disc model (\texttt{sim6}; $H/R=0.10$), and lower-viscosity model (\texttt{sim8}; $\alpha=0.15$). Unless explicitly stated in the row label, the remaining parameters are the same as in the fiducial model. The columns show the same evolutionary stages for each model: the initial relaxed state ($t=0$), just prior to the first mid-plane crossing ($t\simeq0.43\,P_0$), and shortly after the second mid-plane crossing ($t\simeq0.72\,P_0$). The green point marks the instantaneous position of the secondary black hole. The colour scale shows the logarithm of the column density. The figures are generated using SPLASH. \label{fig:morphpara}}
\end{figure*}

To systematically assess how the disc response depends on both binary and disc properties, we compare a suite of simulations with varying parameters, as shown in Figure~\ref{fig:morphpara}.
Each row corresponds to a different model, while the columns represent the same evolutionary stages: the initial relaxed state ($t=0$), just prior to the first mid-plane crossing ($t \simeq 0.43\,P_0$), and shortly after the second mid-plane crossing ($t \simeq 0.72\,P_0$).

The first two rows illustrate the dependence on the binary mass ratio. Comparing the fiducial case (\texttt{sim1}; $q=0.0082$) with a higher-mass-ratio case (\texttt{sim3}; $q=0.1$), while keeping all other parameters fixed, shows that increasing $q$ significantly amplifies the disc response. In the low-$q$ case, the perturbation remains weak and localized, producing only mild distortions and faint spiral features, with the disc retaining a largely axisymmetric structure. In contrast, the high-$q$ case exhibits a strongly non-linear response, with prominent spiral arms, enhanced density contrasts, and large-scale asymmetries, indicating that the disc is globally disturbed by the interaction.

The role of orbital eccentricity is illustrated by comparing \texttt{sim3} ($e=0.7$) and \texttt{sim5} ($e=0.9$). Although the orbital evolution proceeds on different timescales, the snapshots correspond to comparable stages of the interaction. In the moderate-eccentricity case, the perturbation develops more gradually, with material sheared along the trajectory of the secondary and forming extended non-axisymmetric features by $t \simeq 0.72\,P_0$. In contrast, the high-eccentricity case exhibits a more impulsive interaction due to the higher crossing velocity, producing stronger shocks and a more violent response. This results in the formation of a compact, high-density structure around the secondary resembling a transient ``pickup'' disc, accompanied by prominent spiral arms. Thus, increasing eccentricity not only strengthens the perturbation but also shifts the interaction toward a more impulsive regime, in which short-lived bound structures form more rapidly and are more clearly identifiable. The implications of this pickup-disc feature are discussed further in Section \ref{pickup} below.

The effect of the disc aspect ratio is shown by comparing \texttt{sim3} ($H/R=0.05$) and \texttt{sim6} ($H/R=0.10$). In the thicker-disc model, the response to the approaching secondary becomes apparent earlier, and even prior to the main crossing, the gas distribution is more extended and disturbed. The larger vertical extent allows the secondary to interact with disc material over a longer path length. Following the crossing, the thinner disc forms a narrower, higher-contrast disturbed structure, with a more compact overdensity near the secondary. In contrast, the thicker disc develops a broader, more diffuse gas concentration and a more extended wake. Thus, increasing $H/R$ modifies not only the strength but also the geometry of the interaction.

Finally, the effect of viscosity is illustrated by comparing \texttt{sim3} ($\alpha=0.26$) and \texttt{sim8} ($\alpha=0.15$), with all other parameters held fixed. In contrast to the variations in mass ratio, eccentricity, or disc aspect ratio, changing the viscosity over this range does not produce a major change in the global morphology of the interaction. In both cases, the secondary crossing generates a similar large-scale non-axisymmetric response, with comparable spiral structure and overall redistribution of gas. The differences are primarily local, with the gas concentration around the secondary remains slightly different. Thus, within the explored parameter range, viscosity mainly influences the fine-scale smoothing of the perturbation rather than the overall morphology of the disc response.

While the morphological response is primarily controlled by the strength and geometry of the perturbation, the key observable consequence is the resulting variability in the mass accretion rate. The perturbations generated during disc crossings redistribute angular momentum and drive gas inward on dynamical timescales, thereby enhancing accretion onto the primary. The efficiency of this inward transport may depend sensitively on both the binary properties and the internal structure of the disc. To assess the robustness and scalability of these enhanced accretion events, we now quantify the amplitude and temporal structure of accretion-rate variability and systematically explore their dependence on system parameters.

\subsection{Mass accretion rate variability}

\begin{figure*}[t!]
	\centering
	\includegraphics[width=1.0\textwidth]{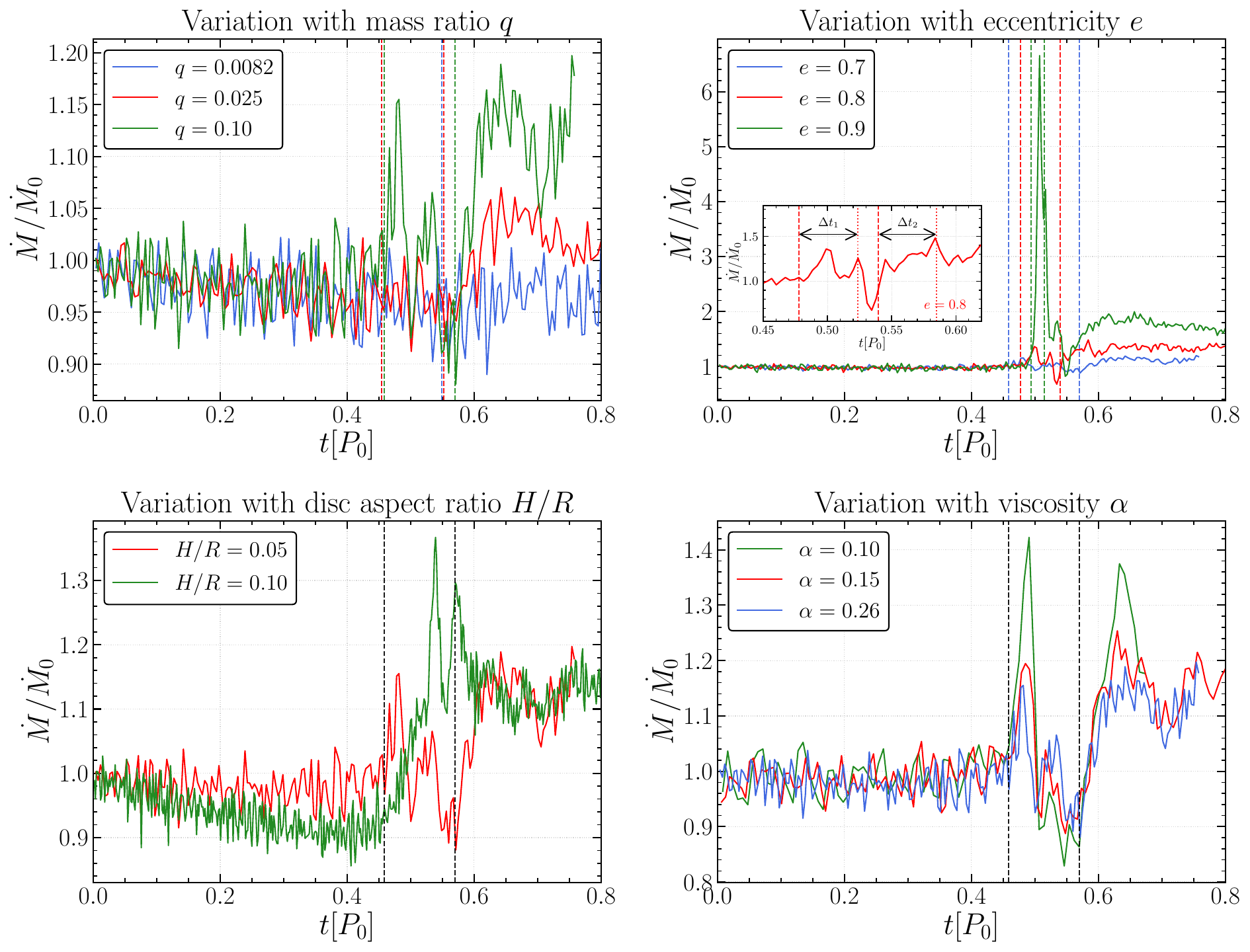}
	\caption{Time evolution of the mass accretion rate onto the primary, normalized by the steady-state value $\dot{M}_0$ for the corresponding disc model. Time is expressed in units of the binary orbital period $P_0 = 2\pi$. The four panels show the dependence on (top-left) mass ratio $q$, (top-right) eccentricity $e$, (bottom-left) disc aspect ratio $H/R$, and (bottom-right) viscosity parameter $\alpha$. The vertical dashed lines indicate the times of the two mid-plane crossings of the secondary, with different colors corresponding to the different parameter values shown in each panel. The inset in the eccentricity panel zooms in on the $e=0.8$ model; the dotted vertical lines mark the corresponding post-crossing accretion peaks, and the horizontal arrows show the measured delays $\Delta t_1$ and $\Delta t_2$ from each crossing to the associated peak. Prior to the crossings, the accretion rate remains close to the steady-state value, while the interaction produces transient enhancements whose amplitude and temporal structure depend on the system parameters. \label{fig:mdot}}
\end{figure*}

We now quantify the dynamical response of the disc through the time-dependent mass accretion rate onto the primary. Figure~\ref{fig:mdot} shows the evolution of the normalized accretion rate, $\dot{M}/\dot{M}_0$, for different binary and disc parameters.
The vertical dashed lines indicate the times of the two mid-plane crossings of the secondary. Prior to the crossings, all models exhibit small fluctuations around unity, reflecting the quasi-steady state of the relaxed disc. Following the crossings, the accretion rate responds with transient enhancements whose amplitudes and temporal structures depend on the system parameters.

The dependence on the binary mass ratio is shown in the top-left panel. Before the disc crossings, the accretion rate in all models fluctuates at the few percent level,
indicating that the disc remains close to its quasi-steady configuration. Following the disc crossings of the secondary, simulations show transient enhancements in the accretion rates. The amplitude of the accretion-rate response increases systematically with the binary mass ratio. In the lowest-mass-ratio case ($q=0.0082$), the accretion rate shows only modest deviations of order $\sim 2$--$4\%$ from the steady-state value. As the mass ratio is increased, the fluctuations from the steady-state value become more pronounced, with peak values reaching around $\sim 5$--$10\%$ above the steady state for $q=0.025$. The strongest response is obtained for the highest mass-ratio case, $q=0.1$, where peak enhancements of order $\sim 15$--$20\%$ are observed. This behaviour reflects the stronger tidal forcing of more massive secondaries, which excite larger-amplitude non-axisymmetric structures and enhance inward mass transport.
A qualitatively similar mass-ratio sensitivity has been found in coplanar circumbinary-disc simulations, where accretion becomes strongly modulated above a transition near $q\simeq0.04$--$0.05$ \citep{2013MNRAS.436.2997D, 2016MNRAS.459.2379D}. More directly relevant to our geometry, recent 3D GRMHD simulations of OJ 287-like systems found that impact-driven spiral shocks and enhanced accretion become weak for $q<0.05$ \citep{2025ApJ...993L..22R}. Although the coplanar circumbinary-disc geometry differs from ours, these studies indicate that sufficiently massive secondaries can drive gas onto unstable, non-axisymmetric trajectories and produce time-dependent accretion.
On the other hand, the timing of the accretion-rate response is broadly similar across all mass ratios, with an enhancement occurring shortly after the disc crossings of the secondary. The timing delay is discussed in detail in Section~\ref{sec:timing}.

The effect of orbital eccentricity is shown in the top-right panel. In contrast to the mass-ratio variation, eccentricity strongly affects both the amplitude and the temporal character of the response. For $e=0.7$, the accretion rate shows only modest enhancement following the crossings. As the eccentricity is increased to $e=0.8$, the accretion response becomes noticeably stronger, with the most extreme behaviour observed for the highest eccentricity case, $e=0.9$. For $e=0.9$, the accretion rate rises rapidly to roughly $6-7$ times the steady-state value over a short timescale, producing a narrow peak that is followed by a phase of sustained elevated accretion. Both the peak amplitude and the duration of the enhancement increase systematically with eccentricity. This strong dependence on eccentricity can be understood in terms of the increasing impulsiveness of the disc crossing.

The dependence on the disc aspect ratio is shown in the bottom-left panel. We observe that both the amplitude and temporal extent of the response depend sensitively on the disc thickness. For the thinnest disc, $H/R=0.05$, the fluctuations in accretion rate remain close to the steady-state level. As the disc aspect ratio is increased to $H/R=0.10$, the accretion response becomes more pronounced, with a clearer and more sustained enhancement following the interaction. This trend reflects the role of pressure support in mediating the disc response to the perturbation.
In our inclined disc-crossing geometry, one possible explanation is that thicker discs present a larger vertical interaction region to the secondary. As a result, the perturber interacts with the gas over a longer time and over a larger vertical extent, allowing the local impulse to couple more effectively to global non-axisymmetric motions. The larger sound speed may further help the disturbance spread into a broader spiral structure, producing a more sustained accretion response.

Finally, the effect of viscosity is shown in the bottom-right panel. We observe transient enhancement in the accretion rate, indicating additional inward mass transport across a range of viscous regimes. We find that discs with higher viscosity exhibit smaller relative peak enhancements in the accretion rate. In contrast, lower-viscosity discs display more impulsive behaviour, characterized by sharper and more localized accretion rate enhancement following the interaction. While viscosity helps angular momentum transport and sets the steady-state accretion rate, it also damps non-axisymmetric perturbations. The timing of the accretion response is largely unaffected by viscosity, indicating that the onset of the inflow event is primarily controlled by the secondary's orbital dynamics, while viscosity determines the response's efficiency and duration.

Taken together, these simulation results show that the amplitude of the accretion-rate variability is primarily governed by the binary parameters. Increasing $q$ enhances the strength of the tidal forcing, while higher $e$ (fixing $a$ and thus larger $e$ means smaller pericenter distance) leads to more impulsive interactions, both of which result in significantly larger accretion-rate peaks. In contrast, the disc parameters ($H/R$ and $\alpha$) mainly regulate how the perturbation is processed, controlling the duration, smoothness, and temporal extent of the response rather than its peak amplitude. In addition to the amplitude, the timing of the accretion response relative to the disc crossings shows a finite delay after the secondary passes through the disc crossings, indicating that the inward transport of gas is not instantaneous but is mediated by the propagation of the disturbance through the disc. We therefore proceed to quantify this delay and examine how it scales with the system parameters.

\subsection{Delay time of the accretion response} \label{sec:timing}

The timing of the accretion response provides a useful diagnostic of the physical mechanism that transports the perturbed gas inward after the secondary crosses the disc. A key insight comes from the viscosity variations shown in Figure~\ref{fig:mdot}. Although the value of $\alpha$ is varied significantly, the timing of the accretion response remains broadly unchanged -- the onset of the rise, the first post-crossing enhancement, and the subsequent large-scale response all occur at nearly the same times. What changes with $\alpha$ is primarily the amplitude and detailed shape of the response. This suggests that the delay between disc crossing and enhanced accretion is not controlled by viscous transport. This conclusion is further supported by a direct comparison with the viscous time at the crossing radius. For the $e=0.7$ run, using $r_{\rm cross}\simeq 0.56$, $H/R=0.05$, and $\alpha=0.26$, the viscous timescale is $t_\nu \sim 10^2$ in code units. In contrast, the observed delays are much shorter than that. The accretion response, therefore, occurs far too rapidly to be mediated by viscous diffusion.
Another relevant comparison is the propagation of pressure waves through the disc. The corresponding timescale is $t_s \sim R/c_s$, which can be expressed as $t_s \sim (H/R)^{-1}\Omega^{-1}$. For the thin discs considered here with $H/R = 0.05$, this implies $t_s \sim 20\,\Omega^{-1}$. The observed delays are significantly shorter than this, indicating that the accretion response cannot be mediated by pressure-wave propagation across the disc.

The remaining timescale is therefore the local dynamical or free-fall time at the crossing radius, which sets the characteristic time for gas to respond to the gravitational potential once it is strongly perturbed. We note that the free-fall time scales as $t_{\rm ff} \propto r^{3/2}$, and thus depends rather sensitively on the radius at which the disc crossing occurs.

To quantify the temporal response of the disc to the secondary impact, we define a delay time between each disc crossing and the subsequent enhancement in the accretion rate. For a given crossing event, we identify the crossing time $t_{\rm cross}$ from the mid-plane passage of the secondary and measure the delay as
\begin{equation}
	\Delta t_i = t_{{\rm peak}, i} - t_{{\rm cross}, i},
\end{equation}
where $t_{{\rm peak}, i}$ corresponds to the time of the first significant post-crossing peak in $\dot{M}(t)$. In cases where multiple peaks are present (as in the $e=0.7$ and $e=0.8$ runs), we adopt the dominant post-crossing peak as the representative response time, as it corresponds to the main inflow episode rather than a weaker precursor.
This measurement is illustrated in the inset of Figure~\ref{fig:mdot} for the $e=0.8$ model, where the dashed vertical lines mark the two mid-plane crossings, the dotted vertical lines mark the associated accretion peaks, and the horizontal arrows indicate $\Delta t_1$ and $\Delta t_2$.

We apply this procedure to the eccentricity models, where the delay is most clearly measurable. The resulting delays for the two crossings within a single orbit, together with the corresponding crossing radii and local dynamical timescales, are
\begin{equation}
	\begin{array}{l}
		e=0.7:\qquad
		\Delta t_1/P_0 \simeq 0.022,\quad
		\Delta t_2/P_0 \simeq 0.072,\quad \\[2pt]
		\hphantom{e=0.7:\qquad}
		r_{\rm cross} \simeq 0.56,\quad
		t_{\rm ff}/P_0 \sim 0.067,\\[6pt]
		
		e=0.8:\qquad
		\Delta t_1/P_0 \simeq 0.046,\quad
		\Delta t_2/P_0 \simeq 0.045,\quad \\[2pt]
		\hphantom{e=0.8:\qquad}
		r_{\rm cross} \simeq 0.40,\quad
		t_{\rm ff}/P_0 \sim 0.040,\\[6pt]
		
		e=0.9:\qquad
		\Delta t_1/P_0 \simeq 0.014,\quad
		\Delta t_2/P_0 \simeq 0.002,\quad \\[2pt]
		\hphantom{e=0.9:\qquad}
		r_{\rm cross} \simeq 0.21,\quad
		t_{\rm ff}/P_0 \sim 0.016 .
	\end{array}
\end{equation}

The measured delays, expressed in units of the binary orbital period $P_0=2\pi$, are therefore of order the local free-fall time, with $\Delta t \sim t_{\rm ff}$ across the eccentricity simulations. As the eccentricity increases, the crossings occur at smaller radii, the corresponding dynamical time decreases, and the accretion response becomes progressively more prompt, approaching an impulsive limit at high $e$.
This timing is consistent with a picture in which the interaction places a fraction of the perturbed gas on highly eccentric or plunging trajectories, allowing it to reach the inner disc on approximately the local dynamical time rather than through slow viscous effects or pressure-wave propagation.

\subsection{Pickup-disc formation and bound mass evolution} \label{pickup}

In this section we examine how gas can become gravitationally bound to the secondary as it passes through the primary black hole's accretion disc. We refer to the secondary-captured gas as the ``pickup'' mass or disc.
We quantify the mass associated with the pickup disk in order to assess the possibility of a separate emission component associated with the subsequent accretion of gas to the secondary. We also measure the coherence of the angular momentum of the captured gas and find it is indeed better described as a disc rather than as an atmosphere.

\begin{figure*}[t!]
	\centering
	\includegraphics[width=1.0\textwidth]{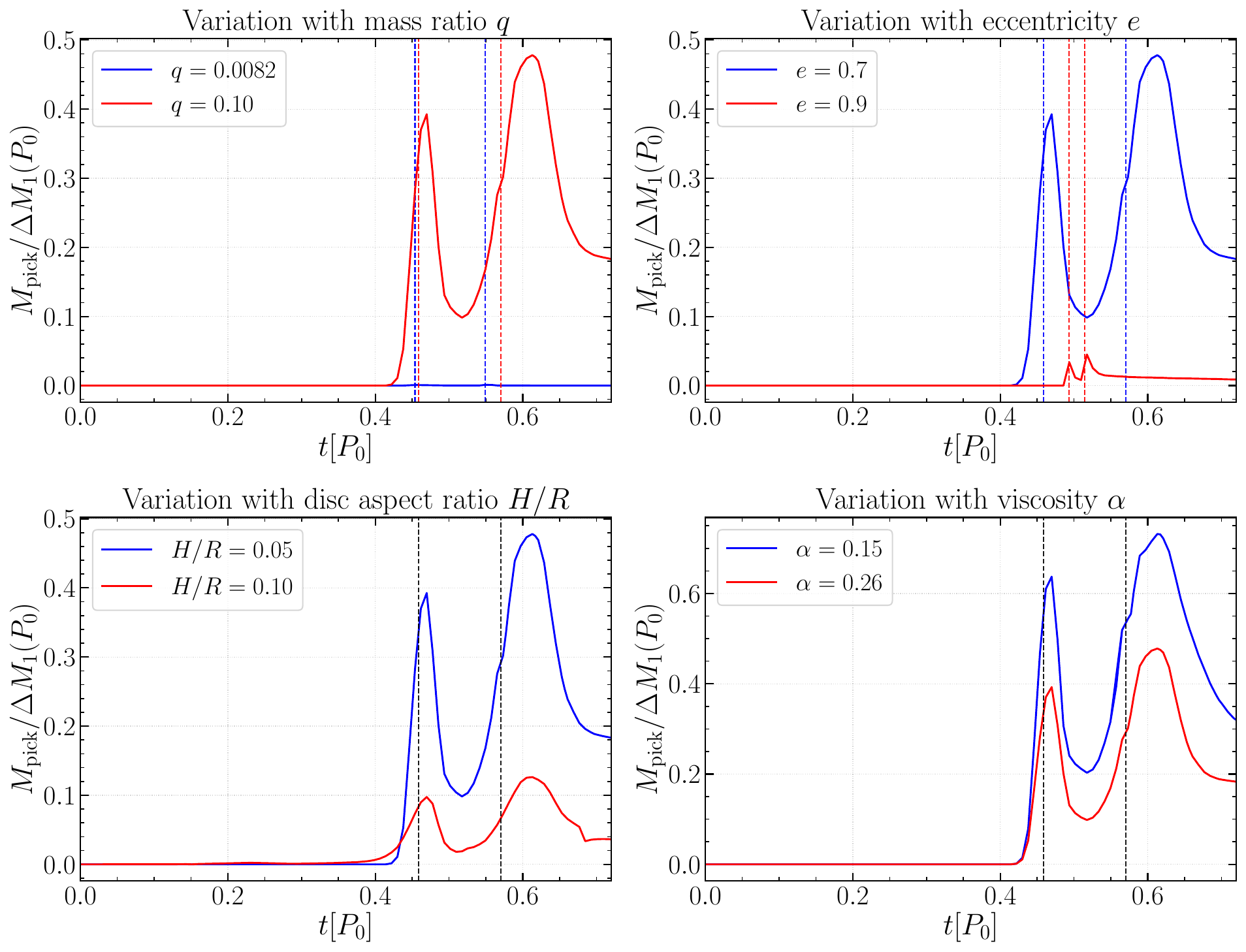}
	\caption{Time evolution of the pickup mass associated with gas bound to the secondary, normalized by the total mass accreted onto the primary over one orbital period, $\Delta M_1(P_0)$. The four panels show the dependence on the binary mass ratio $q$ (top left), eccentricity $e$ (top right), disc aspect ratio $H/R$ (bottom left), and viscosity parameter $\alpha$ (bottom right). The vertical dashed lines mark the two disc-crossing times of the secondary for each model. In all cases, the pickup mass remains negligible prior to the first crossing and rises only after the secondary interacts with the disc. The strongest dependence is observed for the binary parameters $q$ and $e$, while variations with $H/R$ and $\alpha$ are comparatively modest.  \label{fig:pickup}}
\end{figure*}

We estimate the pickup mass by integrating the mass in the simulation that satisfies two criteria. The first is that the particle has negative total specific energy in the frame of the secondary,
\begin{equation}
	\epsilon_{2,i} = \frac{1}{2} |\mathbf{v}_i - \mathbf{v}_2|^2 - \frac{G M_2}{|\mathbf{r}_i - \mathbf{r}_2|},
	\label{eq:relative-energy}
\end{equation}
where $M_2$ is the mass of the secondary black hole, and $\mathbf{r}_i$ and $\mathbf{v}_i$ are the position and velocity of the $i$th gas particle. This criterion by itself is too permissive because it depends on the relative velocity between the gas and the secondary, which becomes small wherever the secondary nearly corotates with the disc. We have found that a substantial surface area of a Keplerian disc can have $\mathbf{v}_i$ close enough to $\mathbf{v}_2$ that the second term in \ref{eq:relative-energy} dominates and disc gas can be spuriously tagged as bound, especially when the secondary is near apocentre of a prograde orbit. We thus impose a second criterion that the mass element lies within the instantaneous Hill radius of the secondary,
\begin{equation}
	R_{\rm Hill} = R_{12}\left(\frac{q}{3}\right)^{1/3},
\end{equation}
where $R_{12}$ is the primary--secondary separation and $q=M_2/M_1$ is the binary mass ratio. We also remark that the Hill-radius criterion alone is too permissive, because during the disc transit considerable mass lies within the Hill radius of the secondary while being not gravitationally bound to it. We have found these two criteria to provide a pickup disc mass estimate compatible with visual inspection and basic expectations, including that zero pickup mass exists in the initial condition.

Figure~\ref{fig:pickup} shows the time evolution of the pickup mass across the explored parameter space. Here, the pickup mass is expressed relative to the nominal mass accreted onto the primary over one orbital period, $\Delta M_1(P_0)$. In all cases, the pickup mass remains negligible prior to the first disc crossing, confirming that gas capture is triggered only after the secondary interacts with the disc. Following the crossing, however, the response depends on the system parameters. The clearest dependence is on the binary mass ratio. As shown in the figure, increasing $q$ leads to a substantial increase in the pickup mass. In the fiducial low-mass-ratio case ($q=0.0082$), the pickup mass remains negligible throughout the evolution. By contrast, for $q=0.1$ the pickup mass rises sharply immediately after the crossings, reaching a pronounced early peak followed by a broader late-time enhancement. This demonstrates that a more massive secondary both perturbs the disc more strongly and has a significantly greater capacity to capture gas. A similarly strong dependence is observed with eccentricity. The $e=0.7$ model produces a substantial pickup mass shortly after the crossings, followed by a stronger late-time enhancement. In contrast, the $e=0.9$ model exhibits a much weaker response. Thus, decreasing eccentricity enhances gas capture by the secondary. Physically, the lower-eccentricity orbit leads to a less impulsive interaction and a longer effective interaction time with the disc, allowing gas to remain bound and accumulate. In contrast, the more rapid passage in the highly eccentric case limits the formation of a substantial circum-secondary reservoir. In contrast to the binary parameters, the dependence on disc properties is comparatively weak. Both $H/R=0.05$ and $H/R=0.10$ produce qualitatively similar pickup-mass histories, with only modest differences in amplitude. Likewise, varying the viscosity between $\alpha=0.15$ and $\alpha=0.26$ leads to broadly similar evolution, with only minor differences in the late-time peak. These findings indicate that, within the explored range, gas capture is governed primarily by the orbital dynamics and tidal influence of the secondary, not by modest variations in disc structure or angular-momentum transport.

As a consistency check, we compare the numerically measured pickup mass with an analytical estimate of the gas initially available within the effective capture cross-section defined in Equation~\ref{eq:cap}. We estimate this mass as $M_{\rm cyl} \simeq \pi R_{\rm cap}^{2}\Sigma(R_{\rm cross})$, where $\Sigma(R_{\rm cross})$ is the disc surface density at the crossing radius. 
Across the $q=0.1$ models, the peak pickup mass is $1.6-2.8M_{\rm cyl}$, while the late-time bound mass is $0.2-1.0M_{\rm cyl}$. For the thin-disc $e=0.7$ runs, the late-time mass agrees with $M_{\rm cyl}$ at the $\sim10\%$ level, with little dependence on $\alpha$. The estimate is less accurate for thicker discs and more eccentric encounters, where a smaller fraction of the initially focused gas remains bound at late times.

To establish whether the pickup mass should be considered a disc, we have measured the total angular momentum $\mathbf{L}_{\rm pick} \equiv \sum_i \mathbf{L}_i$ of the pickup particles with respect to the secondary black hole, where the sum runs over the particles satisfying the two criteria described above. We compared the magnitude of this quantity with the sum of the individual angular-momentum magnitudes through the coherence parameter, $J_{\rm coh} = {\left|\sum_i\mathbf{L}_i\right|}/{\sum_i|\mathbf{L}_i|}$. This quantity approaches unity when the particle angular momenta are aligned. At late times, when the pickup mass became approximately steady, we found $J_{\rm coh}\sim0.9$, indicating that the captured gas is rotationally coherent and is therefore better described as a gravitationally captured disc-like morphology around the secondary, and that it should subsequently accrete viscously rather than as a Bondi-like flow. We note that, because we have chosen a very small sink radius for the secondary (Section~\ref{sec:sinks}), simulation particles are not significantly subtracted in the range of the secondary sink.

\section{Discussion and Conclusions} \label{sec5}

We have investigated the hydrodynamical response of an accretion disc to disc-crossing encounters in eccentric binary black hole systems, motivated by the flare phenomenology of OJ 287.
Recent studies have examined complementary aspects of this problem, including prompt impact flares \citep{2025ApJ...993L..22R}, long-term hydrodynamical evolution \citep{2026ApJ...998..322C}, twisted or warped primary discs \citep{2026arXiv260509517Z}, and magnetic-reconnection-driven emission in OJ 287-like systems \citep{2025A&A...696A..96B}. In contrast, our three-dimensional SPH calculations isolate the response to individual disc-crossing events and quantify how binary and disc parameters regulate tidal transport, delayed accretion variability, and gas capture by the secondary.

We first examined a fiducial model motivated by current constraints on OJ 287 and found that, while disc crossings do perturb the flow,
the resulting enhancement in the black hole mass accretion rate, measured through the primary sink, remains modest.
This motivates a broader exploration of parameter space to determine the conditions under which disc-crossing events produce dynamically significant responses of the inner disk, and in turn modulation of the downstream jet emission.
Our parameter study shows that the efficiency of mass delivery to the primary black hole is mainly controlled by the binary mass ratio, with substantial delivery occurring only for relatively massive perturbers, $q \gtrsim 0.1$.
Increasing eccentricity changes the character of the response by making the crossings more impulsive, shifting the interaction to smaller radii, and producing sharp peaks in the black hole mass accretion rate.
Variations in disc thickness and viscosity similarly influence the detailed structure of the response, but their effects remain secondary compared to the dominant role of the binary mass ratio.

A key result is that the enhancement in the primary black hole mass accretion rate lags the disc crossing by approximately the free-fall time from the impact radius. We can rule out a viscous delay because varying $\alpha$ changes the amplitude of the response (with smaller $\alpha$ causing modestly larger gas delivery), but does not influence the time of the delivery, as would be expected if the inflow time were viscous. The delay instead indicates a dynamical response, in which a fraction of the perturbed gas, or the associated shock-driven disturbance, propagates inward quasi-ballistically. For an OJ 287-like scaling with $P_{\rm orb}\simeq 12\,{\rm yr}$, the measured delays of $t_{\rm ff}/P_0\sim 0.016$--$0.067$ correspond to physical delays of order a few months to $\sim 1$ yr between the disc-crossing and the subsequent enhancement in primary black hole accretion. If this accretion enhancement modulates the jet power, the observed jet variability may be further delayed by the propagation of the disturbance to the downstream jet emission zone.

A further important result is the formation of transient circum-secondary, or ``pickup'', discs during sufficiently strong encounters. Gas capture is enhanced for more massive perturbers and for lower-eccentricity, less impulsive encounters. Consistent with this trend, the late-time pickup mass in the thin disc, $e=0.7$ models agrees to within about 10\% with the gas initially contained within a capture radius set by the smaller of the Bondi--Hoyle--Lyttleton and Hill radii. This seems to validate assumptions that have been adopted in other works for gas capture onto compact objects interacting with AGN discs \citep{2024MNRAS.531.4656W,2025MNRAS.543.3768W,2026arXiv260300226L}.
For OJ 287-like parameters, and assuming the primary accretes at a rate comparable to the Eddington limit, this implies a pickup-disc mass of order $10^2$--$10^3\,\rm{M}_\odot$ is possible. If even a fraction of this material subsequently accretes over a timescale comparable to or smaller than the orbital period, the corresponding accretion rate onto the secondary can approach the Eddington limit, yielding luminosities $\gtrsim 10^{46}$--$10^{47}\,{\rm erg\,s^{-1}}$. For a supermassive secondary with mass $M_2 \sim 10^9\,\rm{M}_\odot$, the characteristic temperature of a standard accretion flow is $T \sim 10^5$ K, implying that the thermal emission would peak in the UV/EUV band.
While the detailed spectral properties depend on the structure and radiative efficiency of the flow, these estimates suggest that the secondary can gravitationally capture enough gas to power a long-lived transient, potentially extending into the UV or soft X-ray bands, that may appear as an additional component of the OJ 287 emission in the months following at least the \emph{upward} disc crossings. We note that after \emph{downward} disc crossings emission from the secondary accretion would be eclipsed by the primary disc if it is optically thick. Observation of an emission component associated with secondary accretion after a downward crossing would imply the primary disk is optically thin.

Our results provide insight into recent observational constraints on OJ 287. In particular, the MOMO non-detection of the predicted 2022 flare \citep{2023MNRAS.522L..84K}, together with the low inferred disc luminosity, favours a primary substantially less massive than that adopted in the ultramassive precessing-binary model \citep{1996ApJ...460..207L, 2018ApJ...866...11D, 2023MNRAS.525.1153V}. At the same time, several OJ 287 outbursts are polarized and non-thermal \citep{2017Galax...5...83V}. The disc-crossing explanation for these events requires the interaction to enhance the gas supply to the jet-launching region. Our simulations independently show that such a strong accretion enhancement requires a larger binary mass ratio, which can be achieved with a less massive primary or a more massive secondary, while the delay remains nearly unchanged. Our interpretation of a lower massive primary is therefore consistent with the MOMO estimate of $M_1 = 1.3\times10^8 \rm{M}_\odot$ from broad-line-region scaling, as well as with the $M_1 \simeq (3.2-4.6)\times 10^8 \rm{M}_\odot$ estimate from host-galaxy bulge-luminosity and velocity-dispersion relations \citep{2002A&A...388L..48L} and the $\sim 10^8 \rm{M}_\odot$ total mass adopted in the jet-precession modelling \citep{2018MNRAS.478.3199B}.

The physics of any emission arising from secondary accretion will require further study. \citet{2023Galax..11...82V} suggested the one-day optical flare in November 2021 to be activity of the secondary jet, while \citet{2024MNRAS.535.3732K} modeled the 2016--2017 Swift/XRT spectra with a bulk-motion Comptonization model and argued that the accreting object is likely the secondary. Nevertheless, neither event seems to uniquely require accretion onto the secondary. 
Our simulations quantify the gas capture by the secondary. For a sufficiently massive perturber, the subsequent accretion could generate an observable UV or soft X-ray emission. Monitoring after disc crossing can therefore constrain the fraction of captured gas that reaches the secondary and associated accretion timescale. Using the corresponding scaling, the characteristic temperature is $T \sim 10^5$ K, implying the thermal peak at $E\sim 30-40$ eV.  
This band is not directly covered by the instruments currently used to monitor OJ 287, it lies above the Swift Ultraviolet/Optical band and below the Swift X-Ray band \citep{2005SSRv..120...95R,2005SSRv..120..165B}. Therefore, with current facilities, secondary accretion must be tested by looking for the observable UV tail of this component, or for additional soft X-ray emission produced by Comptonization of the disc photons \citep{2012MNRAS.420.1848D,2018A&A...611A..59P}. On the other hand, non-detection of an additional UV or soft X-ray component cannot exclude a pickup disc, unless a long viscous accretion time exceeding the orbital time could also be excluded.

\section*{acknowledgments}
DG acknowledges financial support from Clemson University's CU Fellows program, and Clemson's Palmetto compute cluster which provided all computational resources for this work. DG is also grateful to Jeffrey Fung for insightful discussions that helped shape this study. JZ acknowledges financial support from the National Science Foundation under Grant Number AST-2408034.

\section*{Data Availability}

The data underlying this article will be shared on reasonable request to the corresponding author.

\bibliography{sample701}{}
\bibliographystyle{aasjournalv7}



\end{document}